\documentclass{ws-mpla}

\begin{document}

\catchline{}{}{}{}{}
%
%%%%%%%%%%%%%%%%%%%%%%%%%%%%%%%%%%%%%%%%%%%%%%%%%%%%%%%%%%%%%%%%%%%%

\title{Lattice QCD at Finite Temperature and Density}
\author{\footnotesize M.P. LOMBARDO}
\address{Laboratori di Frascati, Istituto Nazionale di Fisica Nucleare
, Via E. fermi 40, I-00044 Frascati(RM) Italy 
lombardo@lnf.infn.it}
\maketitle

\pub{Received (Day Month Year)}{Revised (Day Month Year)}

\begin{abstract}
A general introduction into the subject
aimed at a general theoretical physics audience. We introduce 
the sign problem posed by finite density lattice QCD, and
we discuss  the main methods proposed to circumvent it, with emphasis
on the imaginary chemical potential approach. The interrelation 
between Taylor
expansion and analytic continuation from imaginary chemical
potential is discussed in detail. The main
applications to the calculation of the critical line, 
and to the thermodynamics of the
hot and normal phase are reviewed..

\end{abstract}

\ccode{PACS Nos.: 12.38.Aw 12.38.Gc 12.28.Mh. 11.15.Ha 11.10.Wx}

\section{Outline}	

This is a general introduction into the subject
aimed at a general theoretical physics audience. 
Several recent  results on the behavior 
of the cold and  hot phases, as well as of the critical line have been 
included, however no attempt was made at being exhaustive: 
further material is to be found in the contributions
by A. Nakamura  and Ph. De Forcrand, as well as in introductory
\cite{intro} and topical reviews \cite{rev}. 

This writeup is organized as follows: we give a first introduction
and physical motivation. Next, we briefly review field theory
thermodynamics, and the lattice approach. Technical difficulties (the sign
problem) are introduced there, while the three main methods to circumvent them
are discussed in the following Section. The following section discusses in some detail the interrelation between the Taylor expansion and the 
imaginary chemical potential approach. Section VI, VII, VIII  give an
overview of the results on the critical line, hot and cold phases respectively.
A short summary concludes the paper.

\section{Introduction}

The historical developments of the phase diagram of CD is characterized
by an increasing complication: early views were based on asymptotic
freedom , and divided sharply the phase diagram into an hadronic phase and 
a quark gluon plasma phase. In the late 90's it has been appreciated that
the high density region is much more complicated than previously thought
\cite{Alford:2003eg}

In the last couple of years, it was the turn of the region 
above $T_c$ to become more 
rich: the survival of bound states above the phase transition 
brought up the idea of a more complicated phase, 
the strongly interactive quark gluon plasma, characterized 
by a rich particle spectrum of colored particle (free quarks,diquarks, etc,.)
\cite {Shuryak:2003ty}.

What has the lattice to say in  this scenario?
Its main task is to provide ab initio calculation of the properties
of the phase diagram, including, of course, the precise location of
the critical  line, and the equilibrium properties of the different phases.
One main challenge, of course, is to frame these results into the more
general context of real time evolution. In this respect, it is worthwhile
to remind ourselves that equilibrium solution
are steady state solution of the dynamical Fokker Planck operator; 
and that lattice calculations will help 
validating simple models which can be studied out of equilibrium:
equilibrium studies are a mandatory first step toward a full
real time understanding of the collisions.

\section{A brief review of field theory thermodynamics }

\subsection{Formulation}
Let us remind ourselves
  how to introduce a chemical 
potential $\mu$
for a conserved charge {$\hat N$} in the density matrix $\hat \rho$
in the Grand Canonical 
formalism, which is
the one appropriate for a relativistic field theory \cite{intro}:
\begin{eqnarray}
\hat \rho &=& e^{-(H - \mu \hat N)/T}   \\ 
\cal Z (T, \mu) & =& Tr \hat \rho = \int d \phi d \psi e^{-S(\phi, \psi)} 
\end{eqnarray}

\subsection{The Hamiltonian Formalism}
My task in this lecture is to discuss numerical results for
the Lagrangian formalism. Still, it is very important to mention
the Hamiltonian approach: early studies did show a great promise, and 
the recent efforts might well indicate that this is correct avenue
to treat the phase diagram at finite temperature and density.
I refer the reader to recent literature on the subject, which contains a 
full set of references to early work \cite{hamiltonian}

\subsection{The Lagrangian Formalism}

The path integral representation of the grand partition function $\cal Z$  
in the Euclidean space gives the temperature as the reciprocal of the
imaginary time:
\begin{equation}
S(\phi, \psi) = \int_0^{1/T} dt \int d^d x {\cal L}(\phi, \psi) 
\end{equation}
with periodic boundary conditions in time for bosons 
$\phi(t=0,\vec x)  = \phi(t = 1/T, \vec x) $
and antiperiodic for fermions 
$ \psi(t=0,\vec x) = - \psi(t = 1/T, \vec x)$.

All in all, ${\cal Z}$ at finite temperature $T$  and density
$\mu$  is the partition function of a statistical system in d+1 dimension, 
where $T$ is the reciprocal of the imaginary time, and $\mu$ couples
to any conserved charge.
This representation, which is the starting point for a lattice
calculation,  allows us to deal with thermodynamics and spectrum
exactly on  the same footing.

The theory is regularized on a space time lattice: a regular 
four dimensional grid with $N_s$ points in each space directions,
$N_t$ points in the imaginary time direction, and spacing $a$.
 We refer to
the very many excellent reviews and textbooks for background material
on lattice field theory,  and
we briefly summarize here
the specific aspects of lattice QCD thermodynamics which will
be useful in the following.

The temperature $T$ on a lattice is the same as in 
the continuum: $T = 1 /N_ta$, $N_ta$ being the lattice extent
in the imaginary time direction (while, ideally, the lattice
spatial size should be infinite).
A lattice realization 
of a finite density of baryons, instead, poses specific problems:
the naive discretization of the continuum expression 
$\mu \bar \psi \gamma_0 \psi$  
would  give an energy
$ \epsilon \propto \frac{\mu^2}{a^2}$ diverging 
in the continuum  (${a \to 0}$) limit \cite{form}.

The problem could be cured by introducing appropriate counterterms, however 
the analogy between $\mu$ and  an external field in the $0_{th}$ 
(temporal) direction offers a nicer solution by
considering the appropriate lattice conserved current \cite{form}. 
This amounts to the following
modification of the fermionic part of
the Lagrangian for the $0_{th}$direction  $L_F^0$:
\begin{equation}
 L_F^0(\mu) = \bar \psi_x \gamma_0 e^{\mu a}\psi_{x + \hat 0} -
      \bar \psi_{x + \hat 0} \gamma_0 e^{-\mu a}\psi_{x }
\end{equation}
while the remaining part of the Lagrangian is unchanged.
This yields the current:
\begin{equation}
J_0 = - \partial_\mu L =- \partial_\mu {L_F}^0(\mu) =
\bar \psi_x \gamma_0 e^{\mu a}\psi_{x + \hat 0} +
      \bar \psi_{x + \hat 0} \gamma_0 e^{-\mu a}\psi_{x } 
\end{equation}
This representation of $J_0$ is amenable  to a simple interpretation: the
time forward propagation is enhanced by $e^{\mu a}$, while
the time backward propagation is discouraged by $e^{-\mu a}$; hence, 
the link formulation generates a  particles--antiparticles asymmetry.
In addition,  note that  $\int J_0 = N - \bar N$ as it should.
An alternative way to look at the link formulation  introduces an
explicit dependence on the fugacity  $e^{\mu/T}$ 
via an unitary transformation for the fields 
\cite{Vink:1988vu}. 
In this way  $ L (\mu) = L (0) $, and the $\mu$ dependence is on 
the boundaries, via the fugacity $e^{\mu/T}$:
$\psi(x + N_T) = -e^{\mu a  N_T} \psi(x) = -e^{\mu/T} \psi(x)$.
This is analogous to the continuum case\cite{evans}.

\subsection{Calculational Schemes}
Having set up the formalism, the task is to  compute
\begin{equation}
{\cal Z}  = \int d U d \psi e^{-S(U, \psi)}
\end{equation}
where from now on the Lagrangian defining the Action will be that of
lattice QCD,
containing gluon fields $U$ and quark fields $\psi$.

We have two options. We might integrate out gluons first: 
\begin{equation}
\int dU d \psi d\bar \psi {\cal Z} (T, \mu, \bar \psi, \psi, U) 
\simeq \int  d \psi d\bar \psi {\cal Z} (T, \mu, \bar \psi, \psi) 
\end{equation}  
This produce an effective {approximate} fermion model:
the procedure is physically appealing, but not systematically improvable,
but for one special (lattice) case (see below).
Alternatively, 
we might integrate out fermions exactly, by taking advantage of the
bilinearity of the fermionic part of the Lagrangian
$L = L_{YM} + L_F = L_{YM} +  \bar \psi M (U) \psi$ :
\begin{equation} 
\int dU d \psi d \bar \psi {\cal Z} (T, \mu, \bar \psi, \psi, U)  = 
\int dU e ^{-(S_{YM}(U) - log(det M))} 
\end{equation}
The ``effective'' model we build this way is exact: the price to
pay being that its physical interpretation is not as clear as for effective
fermion models. Anyway, this expression is the starting point for numerical 
calculations: the fact that in many cases they are highly successful
tell us that the configuration space is well behaved enough that
only a minor subset of configurations, although carefully chosen via 
importance sampling,
suffice to produce reasonable results.

\subsection {Effective Fermionic Models: analytical approaches }

Let us start by following the first idea, namely integrating
out the gluon fields so to define an effective fermionic Action.
This is a time honored approach, leading, for instance, to 
the instanton model Hamiltonian, hence to the exciting
discoveries on the QCD phase diagram of the last five years\cite{effe} .

On the lattice, one very interesting approach leading
to a fermionic model is provided by
the strong coupling expansion: in the infinite gauge coupling limit 
the Yang Mills term decouples from the Action, and the
integral over the gauge fields can be carried out exactly.

 The starting point is the QCD lattice Lagrangian:
 \begin{eqnarray}
 S &=& -1/2 \sum_{x} \sum_{j=1}^3 
  \eta_j(x) [ \bar \chi(x) U_j(x) \chi(x + j) - \bar \chi (x + j)
 U^\dagger_j(x) \chi (x)]  \\
 && -1/2 \sum_x \eta_0(x) [ \bar \chi(x) U_0(x) \chi(x + 0) 
 - \bar \chi (x + 0) U^\dagger_0(x) \chi (x)]  \nonumber \\
 && -1/3 \sum_{x} 6/g^2 \sum_{\mu,\nu=1}^4[ 1 - re Tr U_{\mu \nu}(x)] 
 \nonumber \\
 && + \sum_x m \bar \chi \chi \nonumber
 \end{eqnarray}
 
 The $\chi, \bar \chi$ are the staggered fermion fields living on the
 lattice sites, the $U$'s are the $SU(N_c)$ gauge connections on the links, the
$\eta$'s are the lattice Kogut--Susskind
 counterparts of the Dirac matrices, and the chemical
 potential is introduced via the time link terms $e^\mu$, $e^{-\mu}$ 
as discussed above. This time
we have written down explicitly the lattice Action to show that the
 pure gauge term \\
 $ S_G = -1/3 \sum_{x} 6/g^2 \sum_{\mu,\nu=1^4 }[ 1 - re Tr U_{\mu \nu}(x)]$
 contains the gauge coupling in the denominator, hence it disappears in
 the infinite coupling limit. Consequently, one can perform independent
 spatial link integrations, leading to
 \begin{equation}
 {\cal Z} = \int \prod_{time links} dU_t 
 d \bar \chi d \chi e ^{-1/{4N} \sum_{ <x,y>}
 \bar \chi (x) \chi (x) \bar \chi (y) \chi (y) }
 e^{-S_t}
 \end{equation}
 where $\sum_{<x,y>}$ means sum over nearest neighboring links, terms
 of higher order have been dropped, and we recognize a four fermion
interaction \cite{strong}. Further manipulations yield 
the mean field  effective potential:
\begin{equation}
\nonumber V_{eff}(<\bar \psi \psi>, {\mu}) =  2 {cosh(rN_tN_c\mu)} + 
   sinh[(N_t+1)N_c <\bar \psi \psi>] 
/sinh(N_t < \bar \psi \psi>) 
\label{eq:sc}
\end{equation}
which we quote for further reference.
A standard  analysis of $V_{eff}$ finally gives the condensate as a function
of temperature and density, and allows the reconstruction of the
phase diagram. 

More recently this approach has been furthered both 
in two \cite{Nishida:2003uj}
and three colors\cite{Bringoltz:2002ug}, and
new developments on cluster algorithms have appeared as well 
\cite{Chandrasekharan:1999cm}. 

In order
to describe in detail the rich physics of the
finite density phase, one needs both to include
higher order terms into the strong coupling expansion, as well
as to go beyond a simple mean field analysis, which assumes an homogeneous
background.  
The question is as to whether such improved strong coupling approaches
would be able to generate a four fermion term with the correct
flavor structure as well as order of magnitude,
thus opening the possibility of a systematically improvable
approach to finite density QCD, including the study of the superconducting
phase.

\subsection{Effective Gluonic Models: Importance Sampling and the positivity
issue}

Let us write again
\begin{equation}
 {\cal Z}(T, \mu) = 
\int dU e ^{-(S_{YM}(U) - log(\det M))}
\end{equation}

When $\det M > 0$ the functional integral can be evaluated
with statistical methods, sampling the configurations according
to their importance $(S_{YM}(U) - log(\det M))$.
For this to be possible the would-be-measure  ($\det M $) has to be positive.

Let me mention at this point that the factorization method 
\cite{factor} \cite{chr}
might alleviate the problems of complex measures by guiding
the simulations along a sensible path in the phase space.
I will not dwell on this interesting development which is not
really in the scope of an introductory review, but I wish to
call on it the attention of the interested reader, as it really
seems to offer some promise, and has been already tested in 
random matrix models.

In QCD with an even number of flavors, and zero chemical
potential,  standard importance sampling simulations are possible
if $\det M$ is real, which is true if
$M^\dagger = - P M P^{-1}$
where $P$ is any non singular matrix. 
In the most popular lattice fermion formulation this holds:
for Wilson fermions $P = \gamma_5$ and for staggered fermions  $P=I$
(note that this basically expresses a particle--antiparticle symmetry).
We will consider staggered fermions from now on.

Consider now the relationship 
$M^\dagger (\mu_B) = - M (-\mu_B) $
implying  that  reality is lost when $ {\rm Re} \mu \ne 0$:
the reality of the determinant is lost, and with it the possibility
of doing simulations with non zero chemical potential, when we want
to create a particle antiparticle asymmetry. On the other
hand a purely imaginary chemical potential does not spoil the
reality of the determinant: indeed, even if an imaginary chemical
potential can be used to extract information at real chemical
potential, it does not create any real particle--antiparticle
asymmetry and it is natural that the fermion determinant
remains real.

Note that in QCD with two color the determinant remains positive
with nonzero real chemical potential: indeed, in that case quarks and
antiquarks transform under equivalent representation of the
color group and are, essentially, the same particle. Other
important models with a real determinant include finite
density of isospin \cite{Sinclair:2003rm} and four fermion models
\cite{GNlattice}. 

All in all, if we want to extract information useful for QCD at
nonzero baryon density by use of standard MonteCarlo sampling
we will have to use information from the accessible region:
$$
 {\rm Re~~}   \mu =  0, {\rm Im~~}  \mu \le 0
$$

\section{Overview of the  methods}
To begin with, it is useful to think of the theory 
in the $T, \mu^2$ plane. Let us then discuss the phase diagram
from the perspective of analyticity and positivity of the partition
function and of the
determinant. One important consideration to keep in mind:
the Gran Canonical partition function has to 
be positive. It is only the determinant which can change
sign, or even be complex, on single configurations.

Let us consider a mapping 
from  complex $\mu$ to complex $\mu^2$.
Because of the symmetry properties of the theory, this mapping
can be done without loss of generality.
Let us note then that ${\cal Z}(\mu^2)$ is real valued for real  $\mu^2$:
this is a situation familiar from condensed matter: the partition function
is real where the external parameter is real, complex otherwise.

The reality region for the partition function represents states which
are physically accessible. The reality region for the determinant
represents the region which is amenable to an importance
sampling calculation: ${\rm Re}  \mu^2 \le 0$.
The methods which have been applied so far are
\begin{itemize}
\item {\bf $\mu=0$} Derivatives, Reweighting, Expanded reweighting 
\item {\bf $\mu^2 \le 0$} Imaginary chemical potential
\end{itemize}
\begin{figure}
{\epsfig{file= 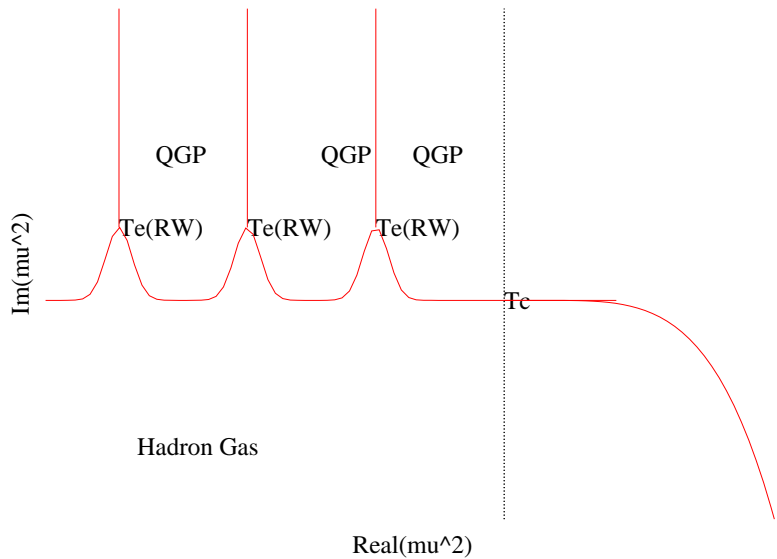, width= 11 truecm}}
\caption{Sketch of the phase diagram in the $\mu^2,T$ plane:
the solid line is the chiral transition, the dashed
line is the Roberge Weiss transition. Simulations can be carried out at
$\mu^2 \le 0$ and results continued to the
physical domain $\mu^2 \ge 0$.
The derivative and reweighting methods have been used so far
to extract informations from simulations performed at $\mu=0.$
The imaginary chemical potential approaches uses results on the
left hand half plane. Different methods could be combined 
to improve the overall performance.}
\end{figure}

\subsection{  Derivatives at $\mu=0.0$}
This is one early attempt at exploring the
physics of nonzero quark density: the derivatives
can be formally computed at $\mu=0$ \cite{deri}. 
The obvious limitation is that we do
not really know how far from the $\mu=0$ axis can we get.
Nonetheless,   such derivatives are interesting per se, and
the region where derivatives are clearly different from zero
is the natural candidate for the application of other
methods.
 
We would also like to quote new results from the Derivatives on
the mass spectrum \cite{Pushkina:2004wa}, where the
first order response of the nucleon (neutron)
to baryochemical potential was computed
in correspondence of several quark masses.

\subsection{ Reweighting from $\mu = 0$ }
Back in the 80's Ian Barbour and collaborators proposed
to calculate 
${\cal Z}(\mu)$ from  simulations at $\mu=0$:
\begin{equation}
{\cal Z} = \left<{{|M(\mu)|}\over {|M(\mu=0)|}}\right>_{\mu=0}
\end{equation}
In other words, the chemical potential $\mu$ of
the  target ensemble at that of the simulation ensemble 
-- $\mu=0$ --  are
different: the properties of the target ensemble can be inferred
from those of the simulation ensemble, provided that there is
a sizable overlap between the two \cite{Barbour:1997ej}.

At $T = 0$ the Glasgow procedure fails because of a poor overlap
(aside, the strong coupling calculations were quite useful to
asses these problems), and it is instructive to study
the overlap problem as seen in the Gross Neveu model, where there
is no sign problem \cite{GNlattice}, 
and the results obtained with reweighting methods
can be compared with those of exact simulations
\cite{Barbour:1999mc}.

The distribution of the order parameter (the $\sigma$ particle)
helps visualizing the problem
the order parameter distributions in the two phases do not 
overlap\cite{GNlattice}.

The conclusion from these early studies was that reweighting fails
in QCD at zero temperature because of a poor overlap, and that
the reason behind the failure is practical rather than conceptual:
the situation can be ameliorated if a better starting point were used.

\subsection{ Fodor and Katz's multiparameter reweighting }
The prescription for ameliorating the overlap is due to 
Fodor and Katz \cite{Fodor:2001au}\cite{Fodor-a}
whose {\em Multiparameter reweighting} 
use fluctuations around $T_c$ at $\mu=0$ to explore the critical region.
Making reference to Fig. 2, and oversimplifying: instead of trying
to reweight the distribution at zero temperature in the broken phase,
which is obviously hopeless, one might hope that a distribution
generated  at zero density, and close to the critical temperature, bears more
resemblance with the target distribution along the critical line,
and is thus amenable to a successful reweighting.

The strategy was applied to QCD \cite{Fodor:2001au}\cite{Fodor-a} .
The improvement obtained is impressive and produced the first
quantitative results for the critical line
at nonzero chemical potential in QCD: we will come back to this
in the section on results. A multistep reweighting proposed
by Crompton \cite{Crompton:2001ws}
might well produce a further improvement.

\subsection{Taylor  Expanded Reweighting}
The Bielefeld-Swansea collaboration suggested a
Taylor expansion of  the reweighting factor as a power series in 
$\lambda=\mu/T$,
and similarly for any operator\cite{Allton:2002zi}
\cite{Allton:2003vx} . 

This strategy is computationally very convenient
as it greatly simplifies the calculation of the determinant.
Expectation values are then given by
\begin{equation}
\langle{\cal O}\rangle_{(\beta,\mu)}=
{
{\langle({\cal O}_0+
 {\cal O}_1\lambda+{\cal O}_2\lambda^2+\ldots)
  \exp({\cal R}_1\lambda+{\cal R}_2\lambda^2+\ldots-\Delta S_g)\rangle_{\lambda=0,\beta_0}}
\over
{\langle\exp({\cal R}_1\lambda+{\cal R}_2\lambda^2+\ldots-\Delta S_g)
\rangle_{\lambda=0,\beta_0}}}.
\end{equation}
Results 
 have been obtained both for the critical line and thermodynamics.

\subsection{Imaginary Baryon Chemical Potential}

This method uses information from all of the negative $\mu^2$ half 
plane (Fig. 1) to explore the positive, physical relevant region.
An imaginary chemical potential $\nu$ in a sense bridges Canonical and
Grand Canonical ensemble\cite{cano}:
\begin{equation}
{\cal Z_C(N)} = \frac {\beta} {2 \pi} \int_0^{2 \pi/ \beta} d \nu {\cal Z_{GC}}
(i \nu) e^{- i \beta \nu N}
\end{equation}
The main physical idea behind any practical application is that 
at $\mu = 0$ fluctuations allow the exploration 
of $N_b \ne 0$ hence tell us about $\mu \ne 0$.
Mutatis mutandis, this is the same condition for the reweighting
methods to be effective: the physics of the simulation
ensemble has to overlap with that of the target ensemble.

A practical way to use the results obtained at negative
$\mu^2$ relies on their analytical continuation in the real
plane. For this to be effective\cite{Lombardo:1999cz}
${\cal Z} (\mu, T) $  must be analytical, nontrivial, and 
fulfilling this rule of thumb:
\begin{equation}
\chi(T,\mu) = \partial \rho (\mu, T) / \partial \mu 
= \partial^2 log Z (\mu, T) / \partial \mu^2
 > 0
\end{equation}

This approach has  been tested
in the strong coupling limit \cite{Lombardo:1999cz} of QCD, in the
dimensionally reduced model of high temperature QCD
\cite{Hart:2000ef}  and, more recently,   in the two
color model \cite{Giudice}.

\section{Taylor expansion and analytic continuation}

We discuss here in more detail the interrelation between the
Taylor expansion and the imaginary chemical potential approach.

Results obtained at imaginary $\mu_B$ can be analytically continued to
real $\mu_B$\cite{rev,Lombardo:1999cz,deForcrand:2002ci,D'Elia:2002gd}. 
In principle,  rigorous arguments guarantee that the 
analytic continuation 
of a function can be  done within the entire analytic domain.
In practice, the exact analytic form is not known, and a systematic procedure
relying on the Taylor expansion is only valid within the circle of
convergence of the series itself.
Here,  we discuss how to implement
the analytic continuation of the critical line and
of thermodynamics observables beyond the circle of convergence
of the Taylor series in a controlled way.

Let us remind ourselves that an analytic function is locally
      representable as  a Taylor series. The convergence  disks can be
      chosen is such a way that they overlap two by two, and cover the
      analytic   domain.  Thus,   one  way   to  build   the  analytic
      continuation   is  by  connecting   all  of   these  convergence
      disks. The arcs of the  convergence circles which are within the
      region where f is analytic have a pure geometric meaning, and by
      no means are an obstacle to the analytic continuation.
      Assume now that the  circle of convergence about $z$  = (0,0) has
      radius  unit, i.e.  is  tangent  to the  lines  which limit  the
      analytic domain; take now a $z$ value, say $z_1 = (0,a), 1/2 < a < 1  $
      inside the convergence  disk as  the origin  of a  new series
      expansion, which is explicitly defined by the rearrangement 
                  $(z- z_0)^n = (z - z_1 + z_1 - z_0)^n$ 
      As the  radius of  convergence of the  new series will  be again
      one, this procedure will extend  the domain of definition of our
      original  function (the  two series  define restrictions  of the
      same function to the intersection between the two disks), and by
      'sliding'  the convergence disk  we can  cover all  the analytic
      strip.

We have skecthed above the standard theoretical argument
to demonstrate the feasibility of analytic continuation beyond the
radius of convergence, and we will show that 
the Pade' series is one practical way to accomplish it.

To complete this discussion, let me mention 
that the radius of convergence of a Taylor
expansion about the origin might well be larger than the distance 
of the origin itself from the
nearest singularity. While complex analysis textbooks offer a full discussion
of this point, I would like to single out here three cases which might be
encountered in usual critical behavior:
$f1(z) = A1(z) (1 - z/z_c)^{-\lambda} ;
f2(z) =  A2(z) \theta(z -z_c) ;
f3(z)  =  A3(z) \theta(z -z_c)(1 - z/z^*)^{-\lambda}$ , where $An(z)$
is an analytic function.

Case 1 corresponds to an usual critical behavior (second order or larger).
Case 2 represents a strong first order phase transition. Case 3 is
intermediate between the two, a weak first order transition at $z_c$,
and a spinodal point at $z^*$. 
\footnote{The analytic continuation is insensitive to
a discontinuous phase transition since it lives on the
metastable branch : it follows the secondary minimum and
determines the spinodal point $<\bar \psi \psi> = A(\mu - \mu^{*})^{\beta}$.
The discontinuity can be related to $\mu - \mu^*$.
Both shrinks to zero at the endpoint of a first order transition.}
Correspondingly, we have  
different radius of convergence of the  Taylor series: 
$r_1 = |z_c| , r_2 = \infty, r_3 = z^*$ : in conclusion
the radius of convergence of the Taylor expansion
for the critical line and thermodynamics observables 
 might be  infinite as well a finite, depending on the nature of the 
Roberge Weiss transition. Conversely, if the nature of the phase transition 
is known, one can infer from it the radius of convergence of the Taylor
series, as done by Gavai and Gupta \cite{Gavai:2004sd}, which in turn
locates the critical point.

\section{ The Critical Line : Taylor vs Pade'}

A full discussion of the current status of the critical line
we refer to recent reviews\cite{rev}. Here, we would like to emphasize more
general aspects on the possibility of continuing the results beyond 
the radius of convergence
of the Taylor expansio, by taking adantage of the discussions
preented in the Section above.

The radius of convergence of the Taylor representation of
the critical line might well be limited by the Roberge Weiss singularities
(see again Fig.1). However, as explained before, the Pade' appoximation is not.

We \cite{lat05} have performed  present the Pade' analysis where we 
have used data  for
four \cite {D'Elia:2002gd} and two flavor \cite{deForcrand:2002ci}.
The results seem stable beyond $\mu_B = 500 MeV (\mu_B/T \simeq 1)$, 
with the Pade' analysis  in good agreement with Taylor expansion
for smaller $\mu$ values. At larger $\mu$ the Taylor expansion 
seems less stable, while the Pade' still converges, giving a slope 
of the critical line larger than the naive continuation of the
second order Taylor approximations. The same behavior is suggested by
recent results within the canonical approach\cite {kra} 
and the DOS method \cite{chr}.

We underscore that the possibility of analytically continue the
results beyond the radius of convergence of the Taylor series
by no means imply that one can blindly extrapolate a lower order
approximation! Even when it is \underline{possible}
to achieve convergence -- via Pade' approximants, or within the
convergence radius of the Taylor series --  one has always to cross 
check different orders of approximation to make sure that convergence has
indeed been achieved. For instance, it would then be interesting to repeat 
the comparisons between the second order  results 
shown in \cite{Azcoiti:2005tv} by extending the Taylor series to
fourth order and/or by use of Pade' approximants.

\subsection{New Results for Wilson Fermions} 

It is very important to double check the results coming from
different methods and different approaches. The results decsribed so
fare have been obtained with staggered fermions. First calculations
with Wilson fermions at fintie density, and imaginary chemical
potential,  have appeared recently{Chen:2004tb,Luo:2004se,Luo:2004mc}. 
The results are so far in nice
agreement with those obtained with the staggered formulation.

\section{ The Hot Phase and the approach to a Free Gas}

The behaviour of the number density at high temperature approaches the lattice
Stephan-Boltzmann prediction, with some residual deviation.
The deviation  from a free field behavior can be parametrised as as 
\cite{Szabo:2003kg,Letessier:2003uj}
\begin{equation}
\Delta P (T, \mu)  =  f(T, \mu) P^L_{free}(T, \mu) 
\end{equation}
where $P^L_{free}(T, \mu)$ is the lattice free result for the pressure.
For instance, in the discussion of Ref. \cite {Letessier:2003uj}
\begin{equation}
f(T, \mu) = 2(1 - 2 \alpha_s/ \pi)
\end{equation}
and the crucial point was that $\alpha_s$ is $\mu$ dependent.

We can search for such a non trivial prefactor $f(T, \mu)$ by taking 
the ratio between the numerical data and the lattice
free field result $ n^L_{free}(\mu_I)$  at imaginary chemical potential:
\begin{equation} 
R(T, \mu_I) = \frac{ n(T,  \mu_I)}{n^L_{free}( \mu_I)}
\end{equation}
A non-trivial (i.e.
not a constant) $R(T, \mu_I)$ would indicate a non-trivial 
$f(T, \mu)$.
In Ref. we calculated $R(T, \mu_I)$ 
versus $\mu_I/T$: the results for $T \ge 1.5 T_c$ seem 
consistent with a free lattice
gas, with an fixed effective number of flavors $N^{eff}_f(T)/ 4 =  R(T) $:
$N^{eff}_f=  0.92 \times 4$ for $T=3.5 T_c$,  and 
$N^{eff}_f = 0.89 \times 4$ for $T = 1.5 T_c$.

\subsection{ Beyond $\mu/T \simeq 1$  in the Hot Phase}

The Pade' approximants to the results for the chiral
condensate in the hot phase have been calculated as well\cite{lat05}
using four flavor data \cite{D'Elia:2004at}.
Again we found that the  Pade' analysis 
seems capable to produce stable results. We should also note that
the Taylor expansion seems stable as well, which might indicate a large 
(infinite?) radius of convergence in this range of temperature.
Indeed, as noted in \cite{D'Elia:2004at} the radius of convergence 
should tend to infinite in the infinite
temperature limit, and indeed it has been estimated to be large 
by the Bielefeld--Swansea collaboration\cite{Allton:2005gk}.
A detailed investigation at imaginary $\mu$ of the region closer to 
the critical temperature is in progress \cite{qcdatwork}. 

We also note a possible 
interplay of thermodynamics and critical behaviour for 
$T_C < T < T_E \simeq 1.1 T_c$ : the critical line at negative
$\mu^2$ would imply , at least for second order and weak first order
transitions, 
$log P(\mu,T)   \propto (\mu^2 - \mu_c^2)^\eta$
which is incompatible with  a free field behaviour.

\section {The Hadronic Phase the Hadron Resonance Gas Model}

The    Hadron Resonance Gas model
might provide as description 
of QCD thermodynamics in the confined, hadronic phase  of QCD:
the grand canonical partition function of the
Hadron Resonance Gas model\cite{Karsch:2003zq} 
has a simple hyperbolic cosine behaviour.

This behaviour could be assessed via a computation of the Taylor 
coefficients \cite{Karsch:2003zq}. As
a perhaps simpler alternative, the hadron gas model
can be framed in the discussion of the phase diagram in the 
temperature-imaginary 
chemical potential plane which suggests
to use Fourier analysis in this region, as observables are periodic
and continuous there\cite{D'Elia:2002gd}.

For observables which are even ($O_e$) or odd ($O_o$) under
$\mu \to -\mu$    the analytic continuation
to real chemical potential of the Fourier series read
$O_e[o](\mu_I, N_t)   =  \sum_n  a_{F}^{(n)} \cosh [\sinh](
n N_t N_c \mu_I)$.
In a Fourier analysis of the chiral condensate
\cite{D'Elia:2002gd}
 and of the number density\cite{D'Elia:2004at} - 
even and odd observables, respectively -  
we limited ourselves to $n=0,1,2$ and we assessed the validity
of the fits via both the value of the $\chi^2/{\rm d.o.f.}$ and the stability
of  $a_{F}^{(0)}$ and $a_{F}^{(1)}$ given by one and two cosine 
[sine] fits:
when HRG holds true, one term in the Fourier 
series should suffice.($sinh (x) \rightarrow sin(x)$)
$n(\mu)  = \frac {\partial P(\mu)}{\partial \mu} = K sin(N_c N_t \mu)$.
 
We found that one cosine [sine] fit  describes reasonably well
the data up to $T \simeq 0.985T_c$; 
further terms in the expansion 
did not modify much the value of
the first coefficients and does not particularly  
improve the $\chi^2/{\rm d.o.f.}$. This means that
our data are well approximated by the hadron resonance gas prediction
$\Delta P \propto (\cosh (\mu_B/T) - 1)$
in the broken phase up to $T \simeq 0.985 T_c$.

In the same region, we can also computer the mismatch with respect to the  HRG
in an 'effective mass analysis' style: this analysis \cite{qcdatwork}
confirms that the HRG is consistent with our data 
within errors up to  $T = .985 T_c$
\begin{equation}
Mismatch = n(\mu) / sin(N_c N_t \mu) - k 
\end{equation}

\subsection { The critical line from the Hadron Gas}

An alternative way to analytically continue the
results relies on phenomenological modeling.
The   Hadron Resonance Gas model
might provide a description 
of QCD thermodynamics in the confined, hadronic phase  of QCD
\cite{Karsch:2003zq,Allton:2005gk,D'Elia:2004at},
and can be used to determine the critical line as well.

The critical temperature as a function of
$\mu_B$ is determined by lines of constant energy
density: $\epsilon \simeq 0.5 - 1.0 $~GeV/fm$^3$\cite {Toublan:2004ks}
A continuation of the critical line using the HRG ansatz
plus a fixed energy (or any other quantity determined at $\mu = 0$ )
criterion suggests the implicit form for the critical line
$
T = f(T) cosh(\mu_B/T)
$ with $\lim_{\mu_B/T \to 0}  f(T) cosh (\mu_B/T) = 1-  k \mu^2$.
We have naively approximated $f(T) =  1-  { k} \mu^2$, and used
the resulting form to fit the data in the $\mu/B < 1$ range.
Accrdoing to the above discussion, this again can be continued
beyond this limit, and   also in this case we get a critical line whose
slope increases with increasing $\mu$\cite{lat05}.

\subsection{How to calculate the critical values}
These calculations described above require the critical
density as an input. Let us then consider:

\begin{equation}
n(i \mu) = a_1 sin(i \mu N_c N_T) + a_2 sin(i 2 \mu N_c N_T)
\end{equation}

Analytic continuation {up to $\mu = \mu_c(T)$) gives:
\begin{equation}
n(\mu)= a_1 sinh (\mu N_c N_t) + a _2 sinh(i 2 \mu N_c N_T)
\end{equation}

This gives the critical density 
at $T = .985T_c$ and mass = $.05$ in lattice units: 
${n_c(\mu_c)/T^3 \simeq 0.5}$ \cite{D'Elia:2004at}, 
once the value of $\mu_c$ has been
taken into account. In addition, 
the mass dependence of the critical density has been be extimated
\cite{D'Elia:2004at} from the Maxwell Relations: \begin{equation}
\partial <\bar \psi \psi> / \partial \mu  = \partial n(\mu) /\partial m
\end{equation}

\section*{Acknowledgments}
I am grateful to the Xiang Luo and the other Organisers of this 
very interesting meeting
for their most kind hospitality and perfect organization. 
This research was supported in part by the U.S. 
National Science Foundation  under Grant No. PHY99-07949.

\section{ Parting Comments}

We have described  three different, independent methods which
afford a quantitative
study of the phases, and phase transitions in QCD. 
All of the methods exploit physical fluctuations either at zero
and purely imaginary chemical potential to explore \underline{real}
baryon chemical potential. The physical idea in a sense is similar, but the
systematics is very different. Cross check are then most useful and
informative, and in many cases, which we have reviewd, have been performed 
satisfactorily.

I should reiterate a caveat: 
both the critical line, and thermodynamics observables in 
the hadronic phase are very sensitive to the quark masses.
I preferred to concentrate on methods and general idea, rather than 
on rapidly evolving results. I apologize for possible incomplete presentation,
and refer once more to the reviews I cited. The reader who chances into
this note at some later stage, might find useful to read the plenary
reviews on Thermodynamics at the Latticexx Conferences and that on
Lattice at the QuarkMatterxx Conferences.  

Finally, again and again, the methods described here are dodges for
the sign problem reviewed above. A more complete solution might well be 
afforded  by the Hamiltonian approach reviewed\cite{hamiltonian}, 
the density of states formalism \cite{chr}, 
the canonical method \cite{kra} \cite{factor}, or the strong coupling
expansion \cite {strong}.


\begin{thebibliography}{99}

\bibitem{intro}
For introductory material,  see e.g. 
S.~Muroya, A.~Nakamura, C.~Nonaka and T.~Takaishi,
Prog.\ Theor.\ Phys.\  {\bf 110}, 615 (2003);
E.~Laermann and O.~Philipsen, 
  Ann.\ Rev.\ Nucl.\ Part.\ Sci.\  {\bf 53} (2003) 16,3
M.~P.~Lombardo,
  Prog.\ Theor.\ Phys.\ Suppl.\  {\bf 153}, 26 (2004)
\bibitem{rev}  S.~Aoki,
  {\em QCD phases in lattice QCD,}
{\tt   arXiv:hep-lat/0509068.} O. Philipsen, plenary talk at Lattice2005,
the XXVI International Symposium on Lattice Field Theory, Dublin, 
22-26 July 2005.
\bibitem{Alford:2003eg}
  M.~Alford,
  %``Dense quark matter in nature,''
  Prog.\ Theor.\ Phys.\ Suppl.\  {\bf 153}, 1 (2004)

\bibitem{Shuryak:2003ty}
  E.~V.~Shuryak and I.~Zahed,
  %``Rethinking the properties of the quark gluon plasma at T approx. T(c),''
  Phys.\ Rev.\ C {\bf 70}, 021901 (2004),

%\cite{Fang:2002rk}
\bibitem{hamiltonian}
  Y.~Z.~Fang and X.~Q.~Luo,
  %``Hamiltonian lattice quantum chromodynamics at finite density with Wilson
  %fermions,''
  Phys.\ Rev.\ D {\bf 69}, 114501 (2004);
Y.~Umino,  Mod.\ Phys.\ Lett.\ A {\bf 17}, 2513 (2002)
  E.~B.~Gregory, S.~H.~Guo, H.~Kroger and X.~Q.~Luo,
  Phys.\ Rev.\ D {\bf 62}, 054508 (2000);
  A.~Le Yaouanc, L.~Oliver, O.~Pene, J.~C.~Raynal, M.~Jarfi and O.~Lazrak,
  Phys.\ Rev.\ D {\bf 38}, 3256 (1988);Phys.\ Rev.\ D {\bf 37}, 3702 (1988).
\bibitem{form}
J.~B.~Kogut, H.~Matsuoka, M.~Stone, H.~W.~Wyld, S.~H.~Shenker, J.~Shigemitsu and D.~K.~Sinclair,
Nucl.\ Phys.\ B {\bf 225}, 93 (1983);
P.~Hasenfratz and F.~Karsch,
Phys.\ Lett.\ B {\bf 125}, 308 (1983).

\bibitem{Vink:1988vu}
P.~E.~Gibbs, Phys.\ Lett. B172, 53(1986); 
J.~C.~Vink,
Nucl.\ Phys.\ B {\bf 323}, 399 (1989).

\bibitem{evans} see 
T.~S.~Evans, {\em The Condensed Matter Limit of Relativistic
QFT} , in 'Dalian Thermal Field' (1985) p. 283, {\tt arXiv:hep-ph/9510298}
for a particularly clear discussion of this point and its implications.

\bibitem{effe}  M.~G.~Alford,
Ann.\ Rev.\ Nucl.\ Part.\ Sci.\  {\bf 51}, 131 (2001); 
R. Rapp, T. Schafer, E.V. Shuryak, M. Velkovsky ,
 Annals Phys. 280 (2000)35.

\bibitem{strong}  P. Damgaard, Phys.Lett. B143, 210 (1984);
van den Doel, Z.Phys. C {\bf29}, 79 (1985) , 
   F.  Ilgenfritz et al., Nucl.Phys. B {\bf377}, 651 (1992); 
    N. Bilic, D. Demeterfi and B. Petersson,  Phys.Rev. D {\bf 37}, 
3691 (1988).

\bibitem{factor}
K.~N.~Anagnostopoulos and J.~Nishimura,
Phys.\ Rev.\ D {\bf 66}, 106008 (2002);
J.~Ambjorn, K.~N.~Anagnostopoulos, J.~Nishimura and J.~J.~M.~Verbaarschot,
JHEP {\bf 0210}, 062 (2002);
V.~Azcoiti, G.~Di Carlo, A.~Galante and V.~Laliena,
Phys.\ Rev.\ Lett.\  {\bf 89}, 141601 (2002);


\bibitem{Nishida:2003uj}
Y.~Nishida, K.~Fukushima and T.~Hatsuda,
  %``Thermodynamics of strong coupling 2-color QCD with chiral and diquark
  %condensates,''
  Phys.\ Rept.\  {\bf 398}, 281 (2004); 
S. Chandrasekharan, talk at Lattice2005.
\bibitem{Bringoltz:2002ug}
B.~Bringoltz and B.~Svetitsky,
Nucl.\ Phys.\ Proc.\ Suppl.\  {\bf 119}, 565 (2003); B.~Bringoltz,
  %``Order from disorder in lattice QCD at high density,''
  Phys.\ Rev.\ D {\bf 69}, 014508 (2004).
\bibitem{Chandrasekharan:1999cm}
S.~Chandrasekharan and U.~J.~Wiese,
Phys.\ Rev.\ Lett.\  {\bf 83}, 3116 (1999).
\bibitem{GNlattice}
S.~Hands, S.~Kim and J.~B.~Kogut,
Nucl.\ Phys.\ B {\bf 442}, 364 (1995); for recent developments see
e.g.  S.~Hands, J.~B.~Kogut, C.~G.~Strouthos and T.~N.~Tran,
Phys.\ Rev.\ D {\bf 68}, 016005 (2003).
\bibitem{Sinclair:2003rm}
D.~K.~Sinclair, J.~B.~Kogut and D.~Toublan,
  %``Finite density lattice gauge theories with positive fermion
  %determinants,''
  Prog.\ Theor.\ Phys.\ Suppl.\  {\bf 153}, 40 (2004)
\bibitem{deri}
C.~Bernard {\it et al.}  [MILC Collaboration],
Nucl.\ Phys.\ Proc.\ Suppl.\  {\bf 119}, 523 (2003);
Physical Review D {\bf 38}, (1988) 2888;
Phys. Rev. Lett {\bf 59} (1987) 2247;
R.V. Gavai and S. Gupta, Phys. Rev. D {\bf 64} (2001) 074506;
Phys. Rev. D {\bf 65} (2002) 094515;
R.V. Gavai, S. Gupta and P. Majumbdar, Phys. Rev. D {\bf 65} (2002) 054506;
Choe  {\it et al}, Physical Review D {\bf 65} (2002) 054501.
\bibitem{Pushkina:2004wa}
  I.~Pushkina {\it et al.}  [QCD-TARO Collaboration],
  %``Properties of hadron screening masses at finite baryonic density,''
  Phys.\ Lett.\ B {\bf 609}, 265 (2005)
\bibitem{Barbour:1997ej}
I.~M.~Barbour,
Nucl.\ Phys.\ Proc.\ Suppl.\  {\bf 26}, 22 (1992);
I.~M.~Barbour, S.~E.~Morrison, E.~G.~Klepfish, J.~B.~Kogut and M.~P.~Lombardo,
Nucl.\ Phys.\ Proc.\ Suppl.\  {\bf 60A}, 220 (1998),

\bibitem{Barbour:1999mc}
I.~Barbour, S.~Hands, J.~B.~Kogut, M.~P.~Lombardo and S.~Morrison,
Nucl.\ Phys.\ B {\bf 557}, 327 (1999).

\bibitem{uno}
N.~Bilic and K.~Demeterfi,
Phys.\ Lett.\ B {\bf 212}, 83 (1988).

\bibitem{Fodor:2001au}
Z.~Fodor and S.~D.~Katz, Phys.\ Lett.\ B {\bf 534}, 87 (2002).

\bibitem{Fodor-a}
Z.~Fodor and S.~D.~Katz, JHEP {\bf 0203}, 014 (2002).

\bibitem{Crompton:2001ws}
P.~R.~Crompton,
Nucl.\ Phys.\ B {\bf 619}, 499 (2001).

\bibitem{Allton:2003vx}
C.~R.~Allton {\em et al.}, Phys.\ Rev.\ D {\bf 68}, 014507 (2003).


\bibitem{Allton:2002zi}
C.~R.~Allton {\it et al.}, Phys.\ Rev.\ D {\bf 66}, 074507 (2002).

\bibitem{Lombardo:1999cz}
M.~P.~Lombardo,
Nucl.\ Phys.\ Proc.\ Suppl.\  {\bf 83}, 375 (2000).

\bibitem{cano}
A.~Hasenfratz and D.~Toussaint,
Nucl.\ Phys.\ B {\bf 371}, 539 (1992);
M. G. Alford, A. Kapustin, F. Wilczek, 
Physical Review {\bf D59} (1999) 054502.


\bibitem{Hart:2000ef}
A.~Hart, M.~Laine and O.~Philipsen,
Phys.\ Lett.\ B {\bf 505}, 141 (2001).

\bibitem{Giudice}
P. Giudice and A. Papa,
  %``Real and imaginary chemical potential in 2-color QCD,''
  Phys.\ Rev.\ D {\bf 69}, 094509 (2004)

\bibitem{deForcrand:2002ci}
Ph.~de Forcrand and O.~Philipsen, Nucl.\ Phys.\ B {\bf 642}, 290 (2002).


\bibitem{D'Elia:2002gd}
M.~D'Elia and M.~P.~Lombardo,
Phys.\ Rev.\ D {\bf 67}, 014505 (2003).


\bibitem{deForcrand:2003hx}
Ph.~de Forcrand and O.~Philipsen,
Nucl.\ Phys.\ B {\bf 673}, 170 (2003).

 \bibitem{Chen:2004tb}
  H.~S.~Chen and X.~Q.~Luo,
  Phys.\ Rev.\ D {\bf 72}, 034504 (2005).

%\cite{Luo:2004se}
\bibitem{Luo:2004se}
  X.~Q.~Luo and H.~S.~Chen,
  %``QCD at finite temperature and density with staggered and Wilson quarks,''
  Nucl.\ Phys.\ Proc.\ Suppl.\  {\bf 140}, 511 (2005).

%\cite{Luo:2004mc}
\bibitem{Luo:2004mc}
  X.~Q.~Luo,
  %``Tricritical point of lattice QCD with Wilson quarks at finite  temperature
  %and density,''
  Phys.\ Rev.\ D {\bf 70}, 091504 (2004).

\bibitem{Karsch:2003zq}
F.~Karsch, K.~Redlich and A.~Tawfik,
Phys.\ Lett.\ B {\bf 571}, 67 (2003).


\bibitem{Fodor:2002km}
Z.~Fodor, S.~D.~Katz and K.~K.~Szabo, Phys.\ Lett.\ B {\bf 568}, 73 (2003).

\bibitem{Letessier:2003uj}
J.~Letessier and J.~Rafelski,
Phys.\ Rev.\ C {\bf 67}, 031902 (2003).
\bibitem{quasi}
F.~Csikor, G.~I.~Egri, Z.~Fodor, S.~D.~Katz, K.~K.~Szabo and A.~I.~Toth,
Prog.\ Theor.\ Phys.\ Suppl.\  {\bf 153}, 93 (2004);
\bibitem {Szabo:2003kg}.~K.~Szabo and A.~I.~Toth,JHEP {\bf 0306}, 008 (2003);
M.~A.~Thaler, R.~A.~Schneider and W.~Weise,
Phys.\ Rev.\ C {\bf 69}, 035210 (2004).
\bibitem{D'Elia:2003uy}
M.~D'Elia and M.~P.~Lombardo,
Phys.\ Rev.\ D {\bf 70}, 074509 (2004).
\bibitem{Hong:2002nn}
D.~K.~Hong and S.~D.~H.~Hsu,
Phys.\ Rev.\ D {\bf 66}, 071501 (2002).

\bibitem{D'Elia:2004at}
  M.~D'Elia and M.~P.~Lombardo,
  %``QCD thermodynamics from an imaginary mu(B): Results on the four flavor
  %lattice model,''
  Phys.\ Rev.\ D {\bf 70}, 074509 (2004)
\bibitem{Azcoiti:2005tv}
V.~Azcoiti, G.~Di Carlo, A.~Galante and V.~Laliena,
  %``Phase diagram of QCD with four quark flavors at finite temperature and
  %baryon density,''
  Nucl.\ Phys.\ B {\bf 723} (2005) 77; V. Laliena, talk at Lattice2005
\bibitem{Gavai:2004sd}
  R.~V.~Gavai and S.~Gupta,
  %``The critical end point of QCD,''
  Phys.\ Rev.\ D {\bf 71} (2005) 114014; S. Gupta, talk at Lattice2005.
\bibitem{Allton:2005gk}
  C.~R.~Allton {\it et al.},
  %``Thermodynamics of two flavor QCD to sixth order in quark chemical
  %potential,''
  Phys.\ Rev.\ D {\bf 71} (2005) 054508 and references therein.

\bibitem{qcdatwork} M. D'Elia, F. Di Renzo, M.P. Lombardo, work
in progress; presented at {\em QCD at work} , Conversano, July 2005.
\bibitem{lat05} M.P. Lombardo, talk at Lattice2005.
\bibitem{kra} Ph. de Forcrand and S. Kratochvila, 
talk at Lattice2005.
\bibitem{chr} C. Schmidt, Z. Fodor, S. Katz, talk at Lattice2005.

\bibitem{Toublan:2004ks}
  D.~Toublan and J.~B.~Kogut,
  %``The QCD phase diagram at nonzero baryon, isospin and strangeness  chemical
  %potentials: Results from a hadron resonance gas model,''
  Phys.\ Lett.\ B {\bf 605} (2005) 129.

\end{thebibliography}
\end{document}